\documentstyle[psfig,prb,aps]{revtex}
\begin{document}
\draft

\twocolumn[\hsize\textwidth\columnwidth\hsize\csname
@twocolumnfalse\endcsname

\title{
Structures and Stabilities of doubly-charged (MgO)$_n$Mg$^{2+}$
(n=1--29) Cluster Ions
}
\author{Andr\'es Aguado$^*$, Francisco L\'opez-Gejo$^{**}$, and Jos\'e M. L\'opez}
\address{Departamento de F\'\i sica Te\'orica, Facultad de Ciencias,
Universidad de Valladolid, 47011 Valladolid, Spain}
\maketitle
\begin{abstract}
{\em Ab initio} perturbed ion plus polarization calculations are reported for
doubly-charged nonstoichiometric (MgO)$_n$Mg$^{2+}$ (n=1--29) cluster ions.
We consider a large number of isomers with full relaxations of the
geometries, and add the correlation correction to the Hartree-Fock
energies for all cluster sizes. The polarization contribution is included at a
semiempirical level also for all cluster sizes. 
Comparison is made with theoretical results for neutral (MgO)$_n$ clusters
and singly-charged alkali-halide cluster ions. Our method is also compared to
phenomenological pair potential models in order to asses their reliability
for calculations on small ionic systems.
The large coordination-dependent polarizabilities of oxide anions favor the
formation of surface sites, and thus bulklike structures begin to dominate 
only after n=24.
The relative stabilities of the cluster ions
against evaporation of a MgO molecule show variations that are
in excellent agreement with the experimental abundance spectra.


\end {abstract}


\vskip2pc]

\section{Introduction}

Cluster science has become a field of intensive research both for 
experimentalists and theoreticians. Small aggregates have a fundamental
interest because they provide a link between the molecular and the solid
state physics, and furthermore they are important in new technological
applications like nanoelectronics. Elucidating the structural and electronic
properties of clusters remains a major challenge for present-day science,
due to the significant deviations that clusters present in their physical and
chemical properties when compared both to the molecule and the bulk.
Another difficulty arises from the huge increase in the number of 
different isomers with
cluster size. Many interesting cluster properties depend
largely on the cluster structure, at least in the case of ionic and
covalently-bonded materials, so a complete description of the
relevant isomer configurations for each cluster size is highly desirable.

In the last years, considerable effort has been devoted to the understanding
of metallic and semiconductor clusters. Meanwhile, studies on metal-oxide
clusters have been comparatively scarce, in despite of their fundamental role
in important physical processes like heterogeneous catalysis. In this work
the interest is focused on magnesium oxide. Stoichiometric MgO clusters have
been investigated both experimental and theoretically. Saunders 
\cite{Sau88,Sau89} reported mass spectra and collision-induced-fragmentation
data for sputtered (MgO)$_n^+$ cluster ions, and Ziemann and Castleman
\cite{Zie91a,Zie91b} performed experimental measurements by using 
laser-ionization time-of-flight mass spectrometry.\cite{Con88,Twu90}
Theoretical calculations have been performed at different levels of accuracy:
simple ionic models based on phenomenological pair potentials
\cite{Mar83,Die84,Phi91} were performed by Ziemann and Castleman \cite{Zie91b}
to explain the global trends found in their experiments; Wilson \cite{Wil97} has
studied neutral (MgO)$_n$ (n$\le$30) clusters by using a compressible-ion
model \cite{Wil96} that includes coordination-dependent oxide
polarizabilities; \cite{Fow88} semiempirical
tight-binding calculations were reported by Moukouri and Noguera;
\cite{Mou92,Mou93} finally, {\em ab initio} calculations on stoichiometric
MgO clusters have been presented recently by Recio {\em et al.},
\cite{Rec93a,Rec93b} Malliavin and Coudray, \cite{Mal97} and de la Puente
{\em et al.} \cite{Pue97} Nonstoichiometric (MgO)$_n$Mg$^+$ and
(MgO)$_n$Mg$^{2+}$ cluster ions have been also detected and studied 
experimentally by Ziemann and Castleman. \cite{Zie91a,Zie91b,Zie91c}
Specifically, they have obtained, by changing the flow rate of the carrier
gas in a gas aggregation source, a mass spectrum comprised almost entirely
of doubly-charged (MgO)$_n$Mg$^{2+}$ cluster ions.\cite{Zie91c} Enhanced
stabilities in the small-size regime were found for n=8,11,13,16,19,22,25,27,
and were explained in terms of compact cubic clusters resembling pieces of the
MgO crystal lattice. If we exclude the pair potential calculations performed
by Ziemann and Castleman, \cite{Zie91b,Zie91c} there is no theoretical
investigation of nonstochiometric MgO cluster ions.

The ``extra'' cation present in nonstoichiometric cluster ions is expected to
result in large structural distortions whenever a specially compact structure
can not be constructed. That is not a problem in pair potential calculations,
where the simplicity of the interactions allows for a complete geometrical
relaxation of the different isomers at a very modest computational cost.
Nevertheless, the need to use some selected set of empirical parameters
for Mg$^{q+}$ and O$^{q-}$ (q=1,2) ions results in a serious questioning of
the reliability of these simple ionic models. On the opposite side, traditional
{\em ab initio} calculations based on the molecular orbital--linear combination
of atomic orbitals approximation are very reliable but 
computationally expensive, the computer time requirements scaling with the
fourth power of the number of atoms in the cluster. Thus, the calculations of
Recio {\em et al.} on (MgO)$_n$ and (MgO)$_n^+$ clusters \cite{Rec93a,Rec93b}
were performed under certain restrictions: (a) the geometries of most of the
isomers were optimized with respect to a single parameter, namely the
nearest-neighbor distance; (b) only 2 or 3 different isomers were considered
for each cluster size; (c) the correlation energy corrections (calculated at
the MP2 level) were included only for n $\le$ 6. In the calculations of
Malliavin and Coudray \cite{Mal97} the geometries were more carefully
optimized but the results were limited to the size range n=1--6.

In the present work we report the results of an extensive and systematic study
of (MgO)$_n$Mg$^{2+}$ cluster ions with n up to 29. We have employed the
{\em ab initio} perturbed ion (aiPI) model, \cite{Lua90a} which is a 
Hartree-Fock model based on the theory of electronic separability 
\cite{Huz71,Huz73} and the {\em ab initio} model potential approach of
Huzinaga {\em et al.}, \cite{Huz87} supplemented with some interaction energy
terms to account for polarization contributions (see next section). The model
has been successfully employed by our group in several studies of alkali
halide clusters and cluster ions. \cite{Ayuel,Agu97a,Agu97b,Ayu98,Agu98}
It has been also used in a study of neutral stoichiometric (MgO)$_n$
(n=1--13) clusters. \cite{Pue97} On one hand, our calculations represent a
major advance with respect to pair potential methods, and on the other hand,
they overcome most of the technical difficulties found in more sophisticated
{\em ab initio} methods: (a) we have allowed for an appropriate geometrical
relaxation of the isomers; (b) we have studied a large set of isomers for
each cluster size (specifically, the total number of isomers studied is around
400); (c) correlation corrections, which have been proved to be essential for an
accurate description of metal-oxide clusters,\cite{Pue97} have been included
for all cluster sizes; (d) we have been able to study relatively large cluster
sizes (up to n=29), enlarging thus the usual size range covered by traditional
{\em ab initio} methods.

The structural results presented in this work could also be useful in the
interpretation of possible future experimental investigations on these
clusters. By measuring the mobility of cluster
ions through an inert buffer gas under the influence of a weak electric field,
drift tube experimental studies
provide valuable information about the cluster 
geometries. \cite{Hel91,Jar95,Mai96} 

The rest of the paper is organized as follows: in section II we give a brief
resume of the aiPI model as applied to clusters (full expositions of the method
have been given elsewhere \cite{Ayuel,Agu97a}), a comparison with
pair potential models which serves to assert the quality of the
methodology, and the details of the computational procedure. The results are
presented in section III, and finally section IV summarizes the main conclusions
extracted from this study.

\section{The aiPI model. Comparison to pair potential models.}

The {\em ab initio} perturbed ion model\cite{Lua90a} was originally designed
for the description of ionic solids,\cite{Lua92} and subsequently adapted to
the study of clusters in our group.\cite{Ayuel,Agu97a,Agu97b,Ayu98,Agu98}
Its theoretical
foundation lies in the theory of electronic separability,\cite{McW94,Fra92}
and its practical implementation in
the Hartree-Fock (HF) version of the theory of electronic
separability.\cite{Huz71,Huz73}
The HF equations of the
cluster are solved stepwise, by breaking the cluster wave
function into local group functions (ionic in nature in our
case). In each iteration, the total energy is minimized with respect to
variations of the electron density localized in a given ion. The electron
densities of the other ions are frozen. In the subsequent iterations each frozen
ion assumes the role of nonfrozen ion.
When the self-consistent process (see more details below) finishes,
the outputs are the total
cluster energy $E_{clus}$ and a set of localized wave functions for
each geometrically nonequivalent ion in the cluster. The cluster energy can be
written as a sum of ionic additive energies:\cite{Ayuel,Agu97a,Agu97b}
\begin{equation}
 E_{clus}=\sum_{R=1}^N E_{add}^R,
\end{equation}
where the sum runs over all ions in the cluster, and the contribution of each
particular ion to the total cluster energy ($E_{add}^R$) can be expressed in
turn as a
sum of intraionic (net) and interionic contributions:
\begin{equation}
 E_{add}^R=E_{net}^R+\frac{1}{2}\sum_{S(\not= R)}E_{int}^{RS}=E_{net}^R+
\frac{1}{2}E_{int}^R.
\end{equation}
The localized nature of the aiPI procedure has some advantages over the usual
molecular orbital models. As the correlation 
energy correction in weakly overlapping systems
is almost intraionic in nature (being therefore a sum of
contributions from each ion), the localized cluster-consistent ionic wave
functions may be used to attain good estimations of this correction. In this 
paper, the correlation energy correction is obtained through Clementi's
Coulomb-Hartree-Fock method.\cite{Cle65,Cha89}
Besides, it also allows the development of computationally efficient codes
\cite{Lua93} which make use of the large multi-zeta basis sets of Clementi and
Roetti\cite{Cle74} for the description of the ions. At this respect, our
optimizations have been performed using basis sets (5s4p) for $Mg^{2+}$
and (5s5p) for $O^{2-}$, respectively. Inclusion of diffuse basis functions
has been checked and shown unnecessary.
Another advantage coming from the localized nature of
the model is the linear scaling of the computational effort with the number of
atoms in the cluster. This has allowed us to study clusters with as many 
as 59 atoms at a reasonable computational cost.

Selfconsistency has been achieved in the following way:
for a given distribution of the ions
forming the cluster, we consider one of them as the active ion R (for
instance, a particular oxygen anion), and solve the Self-Consistent-Field 
equations
for anion R
in the field of the remaining ions, which are considered frozen at this
stage.
Next, we take another
oxygen anion (anion S) as the active ion and repeat the same process.
If the anion S is geometrically inequivalent to anion R, the energy
eigenvalues and wave functions of electrons in anions S are 
different from those of anions R.
We continue this process 
in the same way until all the anions have been
exhausted. 
The same procedure is then followed for the magnesium cations. The process
just described is a PI cycle. We iterate the PI cycles until
convergence in the total energy of the cluster is achieved.
Note that the selfconsistent process can be accelerated if equivalences
between anions or cations are imposed by fixing the symmetry of an isomer. In
order to allow for completely general distortions, we have not employed that
simplification in the present study. Nevertheless,
equivalences in some ionic wave functions have been observed at the
end of the calculations for some highly symmetrical isomers.

It is very interesting to compare any quantum model designed for the study of
ionic materials with the rigid ion and polarizable ion models.
\cite{Mar83,Die84,Phi91} These pair potential models are very intuitive, and
include in a phenomenological way all the relevant terms in the interatomic
potential energy. The quality of an {\em ab initio} method developed for
application on ionic materials can thus be assested by specifying in which
way it improves over a pair potential description.
The binding energy in a pair potential model can be expressed as
$E_{bind}=\frac{1}{2}\sum_{i\not= j}V_{ij} + \sum_iE_i^S$, where $V_{ij}$
is the potential energy between ions i and j, and $E_i^S$ is the self-energy
of the ion i measured relative to the self-energy of the isolated atom.
In the rigid ion model, we have \cite{Phi91}
\begin{equation}
V_{ij}^{rigid}=\frac{q_iq_j}{r_{ij}} + A_{ij}e^{-r_{ij}/\rho},
~~~~~~ E_i^{S,rigid}=0.
\end{equation}
The first term is the electrostatic Coulomb energy between two point ions with
charges q$_i$ and q$_j$, and the second term is the short-range repulsive
Born-Mayer energy reflecting the mutual repulsion due to the overlap of the
wave functions of the ions. In  the polarizable ion model, a polarizability
$\alpha_i$ is assigned to each ion so that the electron shells can be polarized
by the electric field created by the other ions in the cluster. Now we have
\cite{Phi91}
\begin{equation}
V_{ij}^{pol}=V_{ij}^{rigid}+V_{ij}^{MD}+V_{ij}^{DD},
~~~~~~ E_i^{S,pol}=\frac{\mu_i^2}{2\alpha}.
\end{equation}
The self-energy is now different from zero, and the interaction energy contains
two new terms: a monopole-dipole and a dipole-dipole interaction term. In the
Born-Mayer repulsive potential, the distance r$_{ij}$ is replaced by an
effective value r$_{ij}^{eff}$ to take into account the deformation of the
electronic shells upon cluster formation.
An intermediate pair potential model exists in which the repulsive radii of ions
are allowed to deform isotropically under the effects of other ions in the
system, but ionic polarizations arising from the positional displacements of
each shell from its own core are not considered. This is the breathing
shell model,\cite{Mat98} and constitutes an important improvement over the
rigid ion model.
Improvements over the basic polarizable-ion model have also been advanced
by Madden and coworkers. \cite{Mad98}
In the aiPI model, the binding energy can be written as a sum of ionic 
contributions, which are in turn expressed as a sum of deformation and
interaction terms \cite{Ayuel,Agu97a,Agu97b}
\begin{equation}
E_{bind} = \sum_RE_{bind}^R = \sum_R(E_{def}^R + \frac{1}{2}E_{int}^R).
\end{equation}
The interaction energy term is of the form
\begin{equation}
E_{int}^R = \sum_{S \not= R}E_{int}^{RS} = \sum_{S \not= R}
(E_{class}^{RS} + E_{nc}^{RS} + E_X^{RS} + E_{overlap}^{RS}),
\end{equation}
where the different energy contributions are: the classical electrostatic
interaction energy between point-like ions; the correction to this energy
due to the finite extension of the ionic wave functions; the exchange 
interaction energy between the electrons of ion R and those of the other ions
in the cluster; and the overlap repulsive energy contribution. \cite{Fra92}
The deformation energy term $E_{def}^R$ is the self-energy of the ion R
measured relative to the self-energy of the isolated ion. It is an intrinsically
quantum-mechanical many-body term that accounts for the energy change
associated to the compression of the ionic wave functions upon cluster
formation, and incorporates the correlation contribution to the binding energy.
All those terms are
calculated in an {\em ab initio} selfconsistent way. Thus, the improvement
over the classical rigid-ion description is clear. Improvement over a
polarizable-ion description can be questionable in principle due to one basic
assumption used in the actual code, namely spherically symmetric 
electron densities, centered on the nuclei. The electron density clouds of the
ions are thus allowed to distort just isotropically under the effect of the other
ions in the cluster, and indeed the aiPI model could be considered as an
{\em ab initio} breathing shell model. In summary, 
although the aiPI method 
treats quantum-mechanically all the terms present in a breathing shell model, 
plus other many-body terms absent from any classical model, it does not
describe the dipolar terms present in a polarizable-ion model because there
are not induced dipoles. In our previous works on alkali halide clusters,
\cite{Ayuel,Agu97a,Agu97b,Ayu98,Agu98}
inclusion of polarization was not considered essential.
However, the O$^{2-}$ anion has a very deformable density cloud (that is a
large coordination-dependent polarizability), 
and dipolar contributions are expected to be more
important than for halide anions. Relaxing the spherical symmetry assumption
would allow in principle for a proper description of those terms, but many of
the computational advantages of the aiPI model (which have allowed us to 
perform such a detailed study) would be lost. The solution we have chosen is
to include the polarization terms in the selfconsistent process with an
extended polarizable point-ion description. \cite{Wil97}
In such a way, the ``enlarged'' model
obtained improves clearly over all classical descriptions, and can be
considered a benchmark for the pair potential calculations. The price
to be paid is the inclusion of parametrized
coordination-dependent polarizabilities
for the ions.
However, the polarizabilities that are
introduced as parameters in the calculations have been obtained from 
accurate {\em ab initio} calculations, \cite{Fow88}
and thus the reliability of the model is reduced just a little
compared to the most sophisticated
{\em ab initio} methods, whereas
complete geometrical distortions can be considered and 
a large number of isomers studied at less cost. 
Furthermore, the only cases in which doubts related to the energetical
ordering of the isomers arise are those showing near-degenerate isomers,
and we will show that those situations are not frequent.

Now we explain the calculational method used in our study of (MgO)$_n$Mg$^{2+}$
cluster ions. We had no knowledge a priori as to what shapes can adopt these
clusters. So we performed an extensive sampling of the potential energy 
surface by generating a large set of random cluster configurations for each
cluster size n with n=1--13. Except for the smallest sizes (n=1,2), tipically
100,000 configurations were generated for each n. All those random configurations
were fully optimized by using a rigid-ion model. The parameters in equation 3
were those used by Ziemann and Castleman, \cite{Zie91b,Zie91c} although some
tests were also performed with the set of parameters of Bush {\em et al.}
\cite{Bus94} Both sets of parameters led to essentially identical structures,
differing just in minor details. These pair potential calculations could be
performed at a low computational cost and therefore were very useful in
locating reasonable initial guesses for the different minima on the potential energy
surface. Then we took tipically the 15--20 lowest-energy isomers obtained in
the pair potential calculations for each cluster size as input geometries for
the aiPI calculations. We also studied structures obtained by adding or removing
in different ways a molecule from those pair-potential structures, of course
whenever such a structure was not already present in the pair-potential results.
Most of the structures obtained in the range n=1--13 could be classified as
a$\times$b$\times$c, 
where a, b, and c indicate the number of atoms along three perpendicular
edges, so for n$>$13 we directly studied geometries with that formula.
We have considered
also stackings of the different planar structures found for n$\le$13. While
we can not rule out completely the possibility of having overlooked some
important isomer, we believe that we have reduced it considerably.
We have performed aiPI+polarization calculations on all those isomers, without
considering any equivalence between the ions. The optimizations of the
geometries have been performed by using a downhill simplex algorithm.
\cite{Nel65,Wil91} For the oxide polarisabilities we have used the values
given by Wilson. \cite{Wil97} The magnesium polarizability is taken equal to
the bulk value.\cite{Fow85}

We finish this section with a comment about the ionic character of our model.
Although the aiPI model describes the MgO cluster ions in terms of Mg$^{2+}$
and O$^{2-}$ units, we would like to stress that it does not enter in conflict
with any possible assignment of fractional charges to each ion in {\em classical}
models. In fact, the charges used by Ziemann and Castleman \cite{Zie91b,Zie91c}
were +/-1, the charges in the potential of Bush {\em et al.} \cite{Bus94}
are +/-2, and both potentials give very similar results. The difference in
the Coulomb part is compensated by a difference in the repulsive part. In
other words, those charges are just parameters. On the other hand, ionic charges
can be derived from quantum-mechanical calculations following the ideas of
Bader. \cite{Bad90} That has been done recently in aiPI calculations on ionic
solids from which fractional ionic charges have been derived. \cite{Pen97}
Moreover, Wilson \cite{Wil97} and Madden and coworkers \cite{Mad98} have shown that
an extended ionic model including accurate polarization terms can account for
several effects traditionally atributed to ``covalency''.
Bulk MgO is excellently described by the aiPI model \cite{Lua90b}, and
our aiPI plus polarization
results for the MgO molecule bond length and vibrational frecuency are
d= 1.77 \AA and $\omega$= 737 cm$^{-1}$, in good agreement with experiment 
\cite{Hub79} and
{\em ab initio} methods. \cite{Rec93a,Mal97} The appropriateness of the method
for the study of metal oxide clusters is thus assested.

\section{Results}

\subsection{Ground State Structures and Low-lying Isomers.}

In the following, we restrict our discussion to the structural properties of the
lowest-lying isomers. The (MgO)$_n$Mg$^{2+}$ aiPI+polarization structures are
shown in Figure 1. The ground state (GS) and one or two low-lying isomers are
given for each n, except for n$\ge$24 where only one isomer is shown.
A cutoff
of 2.54 \AA, that exceeds by $\approx$ 20 \% the Mg--O distance in the bulk, was
arbitrarily chosen in order to decide whether an O atom is bonded to a Mg
neighbor or not. The energy difference with respect to the most stable isomer
is given (in eV) below each isomer. For cluster sizes n=1 and n=2 (not shown
in the figure), a chain is obtained as the most stable isomer. 
The emergence of the
bulk crystalline structure can be appreciated from Fig. 1: one-dimensional chains
are observed for the smallest cluster sizes (n=1--2);
a planar two-dimensional structure is the GS for n=3;
next there is a size region (n=4--17) in which three-dimensional (3D)
structures are already predominant, but without an establishment of the bulk
symmetry. Specifically, all the ions in those structures are in surface-like
sites. The establishment of face-centered-cubic structures starts at n=18, but
is not complete until n=24. From that value on, all the ground state isomers are
fragments of a rocksalt lattice.

Let us describe now in more detail the structures obtained as a function of n,
first for the small cluster sizes (n$<$17). 
Linear chains are the most stable isomers for
n=1 and n=2. The n=3 GS is a planar rectangular structure with a cation attached
to a corner. A structure obtained by removing an anion from a perfect cube
appears 5.12 eV above, 
and a chain is observed as well as a noncompetitive isomer.
The first three--dimensional ground state isomer
occurs for n=4. It
is obtained by adding a cation to
a (MgO)$_4$ cubic cluster.
A quasiplanar 3$\times$3 sheet and the
chain are not 
competitive anymore. For n=5,
a 2$\times$2$\times$3 piece with an anion removed from a
corner is obtained as the ground state isomer.
From n=5 to n=17, all the ground state isomers are obtained by adding a MgO
molecule either to the GS or to a low-lying isomer of the (MgO)$_{n-1}$Mg$^{2+}$
cluster, in such a way that no bulk ions (ions with coordination six) are
present. The formation of surface sites (with coordination equal to 3, 4,
or 5) is
favored instead. This is a direct consequence of the larger polarizabilities of
oxide anions with lower coordinations. The minimization of just the
electrostatic Coulomb energy between point ions would lead to the formation
of more compact bulklike structures, for which the ions would tend to attain
its full first-coordination sphere. But when polarization effects are included,
the dipole stabilization energy has to be minimized also, favoring a reduction
of the coordination number and the formation of surface sites, the same
conclusion achieved by Wilson \cite{Wil97} in his study of neutral (MgO)$_n$
clusters. As a result that is really worth of mentioning, the GS structure for
n=13 is not a 3$\times$3$\times$3 perfect cube, that is what one would 
have ``expected''
by comparing with the situation encountered for alkali halide cluster ions. \cite{Ayu98,Agu98,Mai96}
This essentially different structural behavior has its roots in the increased
polarizability of the oxide compared to halide anions, and can be accounted
for by an enlarged ionic model like the one used here, without resorting to
the inclusion of ``covalent'' effects. \cite{Wil97,Mad98}
The 3$\times$3$\times$3 (MgO)$_{13}$Mg$^{2+}$ isomer has an oxide anion with
coordination 6, so with a small polarizability. Moreover, the corner sites of that
isomer are occupied by cations, so there is not any oxide anion with
coordination 3. All this results in a small polarization contribution to the
energy, that is not compensated by the increased Madelung energy.

Eventually there has to be a cluster size where
bulklike fragments begin to dominate.
The situation changes a little at n=18,
where a ground state isomer with bulklike ions appears for the
first time, specifically
a 3$\times$3$\times$4 cubic structure with a cation attached to a corner.
Let us note that a major difference between this structure and the 
3$\times$3$\times$3 (MgO)$_{13}$Mg$^{2+}$ isomer is that now there are
oxide anions in corner positions which enhance the polarization contribution.
This enhancement, together with the increased Madelung term associated to a
bulklike fragment, make that isomer more stable than the one obtained by
adding a MgO molecule to the (MgO)$_{17}$Mg$^{2+}$ ground state.
The establishment of bulklike symmetry is not complete yet, however.
For n=19, a 3$\times$3$\times$4 structure with three ions attached is still
favored, even more than for n=18, because now none of the four oxide anions
with coordination 3 is capped by a magnesium. By adding 5 ions to such a
cubic structure, however, at least two oxide anions in corner sites are
capped, and so the GS structure of (MgO)$_{20}$Mg$^{2+}$ is again an isomer
with only surface sites. For n=22, a 3$\times$3$\times$5 perfect structure
can be formed, in analogy with the case of n=13. Such a structure would have
no oxide anions in corner positions
and two oxides with coordination 6, however,
and it is not still energetically favored.
From n=24 on, all the structures obtained are fragments of a bulklike 
crystalline lattice. Although it is possible that for some larger cluster
size the ground state structure will be not yet the most compact face-centered-cubic
fragment, we think we can accept with relative safety n=24 as the critical
size where bulklike symmetry emerges. This is consistent with the
critical size (n=30) estimated by Wilson \cite{Wil97} for neutral (MgO)$_n$ clusters.

Our conclusions suggest substantial structural differences between (MgO)$_n$Mg$^{2+}$
cations and (MgO)$_n$O$^{2-}$ anions. For example, the perfect 3$\times$3$\times$3
cube isomer of (MgO)$_{13}$O$^{2-}$ would not have any oxide anion with 
coordination six, and all the eight corner sites would be occupied by oxides.
The case of 3$\times$3$\times$5 (MgO)$_{22}$O$^{2-}$ isomer would be similar.
We have performed additional calculations for those two cluster sizes and
have obtained the 3$\times$3$\times$3 and 3$\times$3$\times$5 fragments as
ground state isomers of (MgO)$_{13}$O$^{2-}$ and (MgO)$_{22}$O$^{2-}$,
respectively.

A comparison with the results obtained for alkali halide cluster ions
\cite{Ayu98,Agu98} and neutral (MgO)$_n$ clusters \cite{Wil97,Pue97} is illustrative.
Hexagonal prysmatic structures are very common for (MgO)$_n$ clusters,
\cite{Wil97,Pue97} as well as for some neutral alkali halide (AX)$_n$ clusters.
\cite{Ayuel,Agu97a,Agu97b} However, cubic structures are clearly predominant
for (MgO)$_n$Mg$^{2+}$ cluster ions, being this the main effect of
nonstoichiometry and net charge: (a) perfect cubic structures can be built
up with an odd number of ions, but an even number of ions is needed
to construct a perfect hexagonal prysm; (b) for ``defect-like'' structures,
the extra charge is screened more effectively in the cuboid-like isomers.
Examples can be found at n=3, where the hexagonal (MgO)$_3$ isomer \cite{Pue97}
converts into a rectangular structure upon adding Mg$^{2+}$ to lower the
repulsion between cations, and at n=10, where adding a triatomic Mg--O--Mg to
the hexagonal prysm of the neutral (MgO)$_9$ \cite{Pue97}
results in an isomer less stable than that obtained by adding Mg--O--Mg
to a rocksalt piece (second isomer in figure 1). 
Let us now
compare with the results obtained for alkali halide cluster ions. 
\cite{Ayu98,Agu98} 
The main difference comes from the polarization contribution. As stated above,
that is less important in alkali halide materials, for which the establishment
of bulklike symmetry is already complete at n=13.
There is just another minor
difference: centered structures, containing an inner alkali cation
with high coordination, were found frecuently for (NaI)$_n$Na$^+$ and 
(CsI)$_n$Cs$^+$, but such isomers are not observed for (MgO)$_n$Mg$^{2+}$.
Evidently, the repulsion between oxide anions is larger than the repulsion
between halide anions, and allocating six O$^{2-}$ anions around a Mg$^{2+}$
cation in energetically less favorable.

We finish this section with a discussion of the validity of pair potential
models and the effects of the polarization correction on cluster structure.
Comparison of our pair potential calculations with the aiPI results lead us
to the following two important points: (a) pair potential models fail to
reproduce quantitatively the expansion of the interionic distances with cluster
size; (b) the energetical ordering of the different isomers for a given cluster
size is not reliable. On the positive side, the shape of the isomers is
usually quite well reproduced, and the main structural correction obtained with
an aiPI calculation is a global scaling of all the interionic distances. This
last observation supports the use of pair potentials to locate good initial
configurations for ionic clusters, from which {\em ab initio} calculations can
be started. The effects of the polarization correction are sizable for n$<17$. 
The correct determination of the ground states would not be achieved with a
model that did not include polarization corrections, as surface sites would not
be specially favored. Moreover, for the smallest cluster sizes (n$<$8), the
dimensionality of the isomers is not reproduced either.
Specifically, linear chains
are the GS isomers for n=1--3 and planar structures are the most stable
isomers for
n=4--7 in aiPI calculations without polarization. Thus, inclusion of polarization
speeds up the emergence of 3D structures.
The imprecisions associated to the lack of polarization corrections turn
smaller as the cluster size increases and bulklike fragments begin to dominate. 

\subsection{Relative Stabilities. Comparison to Experimental Results.}

In order to study the relative stabilities of (MgO)$_n$Mg$^{2+}$ cluster ions
we calculate the evaporation threshold energy \cite{Phi91,Agu98} required to
remove a MgO molecule from the ground state isomer. For a cluster (MgO)$_n$Mg$^{2+}$
this is done as follows: we consider the optimized GS structure and identify the
MgO molecule that contributes the least to the cluster binding energy.
Then we remove that molecule and relax the resulting 
(MgO)$_{n-1}$Mg$^{2+}$ fragment to the nearest local minimum. The total energy
required to evaporate a molecule is then the difference between the energy of
the parent cluster and the energies of the two fragments, one of which is the
MgO molecule:
\begin{eqnarray}
E_{evaporation}(n) = E_{cluster}[(MgO)_{n-1}Mg^{2+}] \nonumber \\ + E(MgO) -
                     E_{cluster}[(MgO)_nMg^{2+}].
\end{eqnarray}
This process can be termed locally adiabatic because both fragments are allowed
to relax to the local minimum energy configuration after the evaporation. For
some cluster sizes, the fragment of size (n-1) left when a MgO molecule is
removed from (MgO)$_n$Mg$^{2+}$ does not lie on the catchment basin of the
(MgO)$_{n-1}$Mg$^{2+}$ GS isomer, so that the evaporation threshold energies
are larger than the energy differences between adjacent ground states minus
E(MgO) in those cases. The evaporation energies are plotted as a function of n
in figure 2. In the experiments performed by Ziemann and Castleman, 
\cite{Zie91c} the abundances in the mass spectra reflect the relative stabilities
of (MgO)$_n$Mg$^{2+}$ cluster ions against evaporation of a MgO molecule.
Starting from the largest cluster size studied, we can appreciate minima in
figure 2 for n=28, 26, 23, 20, 17, 14, 12, 9, and 5. Thus, evaporation of a
MgO molecule from those clusters is easy relative to evaporation from clusters
of other sizes. Moreover, the curve shows abrupt increases
precisely at those cluster sizes, so that evaporation of a MgO molecule from
the clusters of size (n-1) is significantly more expensive than evaporation
from cluster ions of size n. All of this results in an enrichment of
(MgO)$_n$Mg$^{2+}$ clusters with n=4, 8, 11, 13, 16, 19, 22, 25, and 27, a
result which is in complete agreement with the experiment. \cite{Zie91c}
Nevertheless, there is not a total correspondence between the experimental
magic numbers and maxima in the evaporation energy curve. The experimental
enhanced abundances result from a balance between two main processes: (a)
evaporation from clusters with size (n+1) results in an enrichment of the
(MgO)$_n$Mg$^{2+}$ clusters; (b) evaporation from clusters with size n has the
opposite effect. A quantity that measures the stability of a (MgO)$_n$Mg$^{2+}$
cluster ion relative both to (MgO)$_{n+1}$Mg$^{2+}$ and (MgO)$_{n-1}$Mg$^{2+}$
cluster ions is thus
\begin{equation}
\Delta_2(n) = E_{evaporation}(n+1) - E_{evaporation}(n),
\end{equation}
where both terms in the difference are calculated using eq. (7) as explained
above.
This quantity is plotted as a function of n in Figure 3. Minima are apparent
at n=4, 8, 11, 13, 16, 19, 22, 25, and 27. Thus, the experimental results
concerning enhanced relative stabilities are properly reproduced by our
calculations.

Opposite to the case of alkali halide clusters \cite{Agu97b} and cluster ions,
\cite{Agu98} the magic numbers observed for (MgO)$_n$Mg$^{2+}$ can not be
clearly explained in terms of a compactness argument.
Instead, all the magic numbers observed verify the following: when removing
the least bound MgO molecule from (MgO)$_n$Mg$^{2+}$ (where n is a magic
number), the resulting (MgO)$_{n-1}$Mg$^{2+}$ is never in the catchment area
of the ground state isomer, and thus the evaporation energy is large. 
Let us note that in some cases (n=13,19,25) this is not a direct consequence of
the shape of the ground state isomers involved, but of the asymmetry between
the different facets of those isomers, which make energetically more favorable
to allocate three ``extra'' ions on a different facet than one ``extra'' ion.
Moreover, the magic numbers (n=4,11,13,16,22,25,27) verify a second
property: when removing the least bound MgO molecule from (MgO)$_{n+1}$Mg$^{2+}$
(where n is again a magic number), the resulting (MgO)$_n$Mg$^{2+}$ structure
is either the ground state structure or a low-lying isomer, and thus the
evaporation energy shows a local minimum for those cluster sizes.
Although there is a striking periodicity of three in the magic
numbers observed between n=13 and n=27, our results indicate that there is not
any special structural motif behind that periodicity.

\section{Summary}

The {\em ab initio} perturbed ion model, supplemented with a semiempirical
treatment of dipolar terms, has been employed in order to study
the structural and energetic properties of (MgO)$_n$Mg$^{2+}$ (n=1--29)
cluster ions. We have used a rigid ion model to generate a large set of
initial configurations for each cluster size. Next, those structures have been
optimized without consider any equivalence between different ions, allowing thus
for an appropriate geometrical relaxation. Correlation corrections, which turned
out to be essential for a proper description of (MgO)$_n$ clusters,\cite{Pue97}
have been added to the Hartree-Fock energies for all cluster sizes.
Inclusion of parametrized coordination-dependent polarizabilities 
reduces just a little
the reliability of the model compared to full {\em ab initio} methods, because
those polarizabilities have been themselves extracted from accurate
{\em ab initio} calculations, and
allows to perform such a systematic study with appropriate geometrical
relaxations at a significantly reduced computational cost.
The structural trends of (MgO)$_n$Mg$^{2+}$ cluster ions have been described.
The emergence of the bulk crystalline symmetry proceeds in different
stages, from one-dimensional configurations for n=1--2 to a planar
structure for n=3, and then to three-dimensional structures with a clear
predominance of cubic structures with surface sites for n=4--17.
The critical size at which bulklike rocksalt fragments dominate the spectrum
is n=24.
For (MgO)$_{13}$Mg$^{2+}$ and (MgO)$_{22}$Mg$^{2+}$, the ``expected''
3$\times$3$\times$3 and 3$\times$3$\times$5 rocksalt isomers 
are neither the ground states nor low-lying isomers. This at first sight
surprising result has been rationalized in terms of the importance of the
polarization contribution, and the preference of oxide anions for surfacelike
(specially corner) sites.
A comparison with the
structures obtained for neutral (MgO)$_n$ clusters \cite{Wil97,Pue97} shows that 
the preference for cuboid-like structures is a direct consequence of
nonstoichiometry and net charge. Comparison with the structures obtained for
alkali halide cluster ions \cite{Agu98} reveals substantial differences that can
be explained in terms of the larger polarizabilities of oxide anions compared
to halide anions, and to a less extent in terms of the larger 
repulsion between oxide anions compared to
the repulsion between halide anions. The main effects of the polarization
correction is to speed up the emergence of 3D structures, to lower the
average coordination, and to favor the formation of surface sites. Once
bulklike 
structures are well stablished polarization plays a less important role.

The relative stabilities of (MgO)$_n$Mg$^{2+}$ cluster ions
have been studied by calculating the energy
required to evaporate a MgO molecule from the GS structures
and the first differences between those evaporation energies, that reflect the
stability of a cluster of size n relative to the stabilities of the neighbor
clusters of sizes (n+1) and (n-1). Both sets of calculations predict the
clusters with n=4, 8, 11, 13, 16, 19, 22, 25, and 27 to be specially stable,
a result that is in complete agreement with the experimental enhanced
abundances reported by Ziemann and Castleman. \cite{Zie91c} 
The interpretation of the enhanced stabilities in terms of highly compact
fragments of a face-centered-cubic crystalline lattice is not appropriate for
such small clusters.
Those enhanced
stabilities have been explained in terms of an analysis of the
evaporation process, which involves the explicit consideration of isomer
structures different from the ground states.

Finally, a comparison has been made with the results obtained by using a
phenomenological pair potential model. This comparison has shown that, while
pair potential calculations are really helpful when obtaining initial guesses
for the isomer geometries, the distances and the energetical ordering of the
isomers
obtained with them are generally not reliable.

$\;$

$\;$

The authors gratefully acknowledge Prof. J. D. Gale for providing us with a copy
of reference 42, Prof. A. W. Castleman Jr. for discussing with us the
experimental results in reference 19, and the referee for his valuable 
comments about oxide polarizabilities.
Work supported by DGES (PB95-0720-C02-01) and Junta de
Castilla y Le\'on (VA63/96). A. Aguado is supported by a predoctoral fellowship
from Junta de Castilla y Le\'on.


{\bf Captions of figures}

$\;$

{\bf Figure 1}. Lowest-energy structure and low-lying isomers of
(MgO)$_n$Mg$^{2+}$ cluster ions. Light balls are Mg$^{2+}$ cations and dark
balls are O$^{2-}$ anions. The energy
difference (in eV) with respect to the most stable structure is given below
the corresponding isomers. For n$\ge$24 only the ground state isomer is shown.

$\;$

{\bf Figure 2}.
Evaporation threshold energies required to 
remove a neutral MgO molecule from
(MgO)$_n$Mg$^{2+}$ cluster ions as a function of n. The local minima in the
evaporation energy curve are shown explicitely.

$\;$

{\bf Figure 3}.
Second derivative $\Delta_2(n)$ for (MgO)$_n$Mg$^{2+}$ cluster ions as a
function of the cluster size. Minima identify those cluster sizes with an
enhanced relative stability.


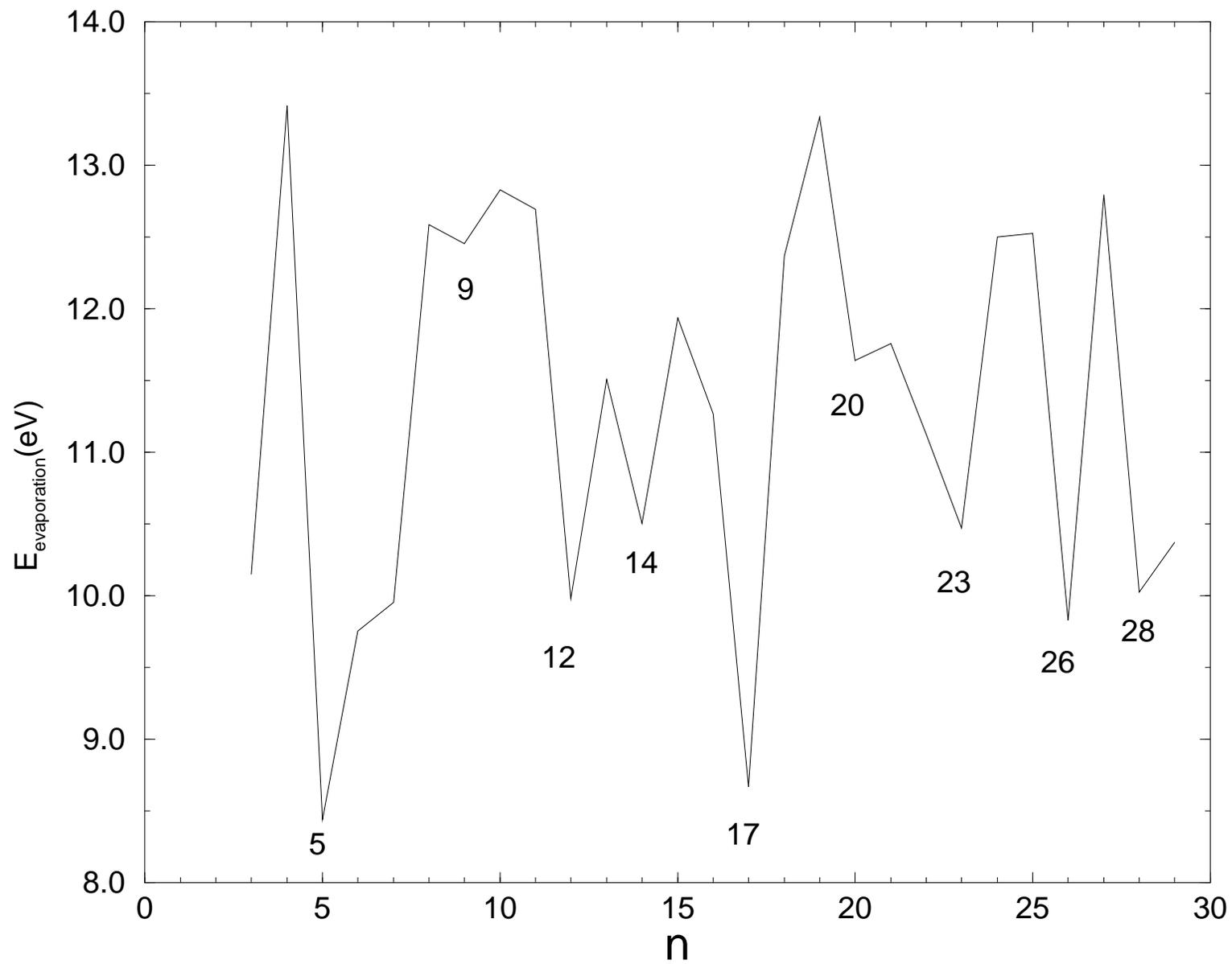
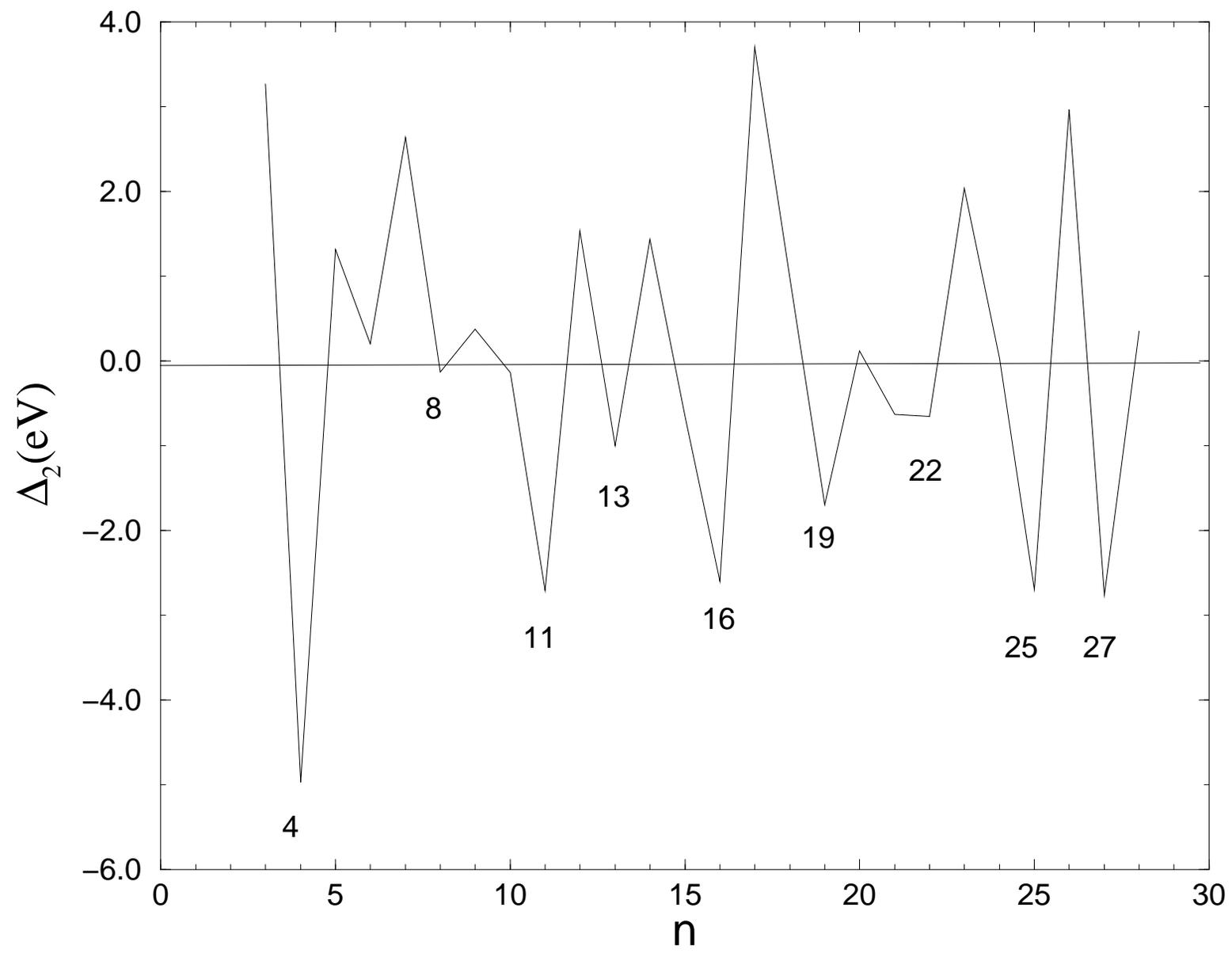
\end{document}